\documentstyle[12pt]{article}
\newcommand{\tr}[1]{\,{\rm tr}\,#1\,}

\begin{document}
\title{
\begin{flushright}
{\small SMI-25-95 }
\end{flushright}
\vspace{2cm}
The Master Field for QCD
\\ and \\ $q$-Deformed Quantum Field Theory}
\author{
I.Ya.Aref'eva
\thanks
{ e-mail: arefeva@arevol.mian.su}
$~$ and $~~$ I.V.Volovich
\thanks{e-mail: volovich@arevol.mian.su}
\\
Steklov Mathematical Institute,\\
Vavilov 42, GSP-1, 117966, Moscow, Russia}
\date {$~$}
\maketitle
\begin {abstract}
The master fields for the large $N$ limit of matrix models and gauge theory
are constructed. The master fields satisfy to standard equations of
relativistic field theory but fields are quantized according to a new rule.
To define the master field we use the Yang-Feldman equation with a free field
quantized in the Boltzmannian Fock space. The master field for gauge theory
does not take values in a finite-dimensional Lie algebra however there is a
non-Abelian gauge symmetry. For the construction of the master field it is
essential to work in Minkowski space-time and to use the Wightman correlation
functions. The BRST quantization of the master field for gauge theory and a
loop equation are considered.
\end {abstract}

\newpage
\section{Introduction}
\setcounter{equation}{0}
There exists an old problem in quantum field theory
how to construct the master field for the large $N$ limit
in QCD. Recently the problem has been reconsidered  \cite {GG}-\cite {AAKV}
by using methods of non-commutative probability theory \cite {Voi,ALV}.

The large $N$ limit in QCD  where $N$ is the number of colours
enables us to understand qualitatively certain striking
phenomenological features of strong interactions \cite {tH}-\cite {AS}.
To perform an analytical investigation one needs to compute the sum of
all planar diagrams. Summation of planar diagrams has been performed only
in low dimensional space-time \cite {BIPZ,QCD2}.

It was suggested  \cite {Wit} that there exists
a master field which dominates the large $N$ limit. The
problem of construction of the master field has been
discussed in many works, see for example \cite {Haan}-\cite {GH}.
Recently Gopakumar and Gross \cite {GG} and Douglas \cite {Doug3}
have constructed the master field for an arbitrary matrix model.
The  construction requires a knowledge of  all correlation functions of a
model. In a previous paper  \cite {AAV} we have presented an  effective
operator realization for the master field (i.e. without knowing an
expression for correlation functions) for a subset of  planar diagrams,
for the so called half-planar diagrams.

The goal of this note is to give an effective operator realization
for the master field for all planar diagrams. We construct the master field
which satisfies to standard equations of relativistic field theory but it
is quantized according to a new rule.

An important fact which we will use is that the master field
for a free matrix field is a new free  field satisfying to unusual
relations $a(k)a^*(k')=\delta(k-k')$ where $a(k)$ and $a^*(k)$ are
annihilation and creation operators.
Therefore an appropriate functional depending on the free matrix field
such as the vacuum expectation of the trace of the product
of the Heisenberg field operators goes in the large $N$ limit
to a corresponding functional on  the new free  field.  To define the master
field we use the Yang-Feldman equation with a free field quantized in the
Boltzmannian Fock space.

For a gauge field $A_{\mu}(x)$ we construct  the master field  $B_{\mu}(x)$.
The field $B_{\mu}(x)$ is similar to the  quantum
photon field in the sense that it does not have matrix indexes. However
the commutator $[B_{\mu}(x),B_{\nu}(x)]$ does not vanish and there is an
infinite dimensional non-Abelian operator gauge symmetry for the gauge
field $B_{\mu}(x)$ with the gauge group being the group of unitary
operators in the Boltzmannian Fock space.

For our construction it is essential to work in Minkowski space-time
and to use the  Wightman correlation functions \cite {SW} and the
Yang-Feldman equation in the Heisenberg representation instead of dealing
with an Euclidean formulation and  using the Feynman correlation
functions and  the interaction representation.

In Section 2 we will consider the master field for a matrix
field model and in Section 3 we will construct the master field
for gauge field. We discuss also the BRST-transformations
for general operator fields (not necessary Lie-algebra valued)
and consider the BRST-quantization of the constructed master field
for gauge theory and  loop equations. We conclude with a discussion  of
a relation of our master field with a  $q$- deformed quantum field theory
and with the Adler generalized quantum dynamics.

\section{Scalar field}
\setcounter{equation}{0}
Let us consider a model of an Hermitian scalar matrix field
$M (x)=(M_{ij} (x)),$ $ i,j=1,...,N$ in the 4-dimensional
Minkowski space-time with the field equations
\begin {equation} 
                                                          \label {1.1}
(\Box + m^{2})M (x)=J(x)
\end   {equation} 
We take the current  $J(x)$ equal to
\begin {equation} 
                                                          \label {1.2}
J(x)=-\frac{g}{N} M^3 (x)
\end   {equation} 
where $g$ is the coupling constant but one can take a more general
polynomial over $M(x)$.
One integrates eq (\ref {1.1}) to get the Yang-Feldman equation
\cite {YF,BD}
\begin {equation} 
                                                          \label {1.3}
M(x)=M^{(in)}(x)+\int D^{ret}(x-y)J(y)dy
\end   {equation} 
where $D^{ret}(x)$ is the retarded Green function for the Klein-Gordon
equation,
$$
D^{ret}(x)=
\frac{1}{(2\pi)^{4}}\int \frac{e^{-ikx}}{m^{2}-k^{2}-
i\epsilon k^{0}}dk
$$
and  $M^{(in)}(x)$ is a  free Bose field. The $U(N)$-invariant
Wightman functions  are defined as
\begin {equation} 
                                                          \label {1.4}
W(x_{1},...,x_{k})=
\frac{1}{N^{1+\frac{k}{2}}}<0|\tr(M(x_{1})...M(x_{k}))|0>
\end   {equation} 
where   $|0>$ is the Fock vacuum for the free field $M^{(in)}(x)$.

We will show that the limit of functions (\ref {1.4})  when
$N\to \infty$ can be expressed in terms of a quantum field $\phi (x)$
(the master field) which is a solution of the equation
\begin {equation} 
                                                          \label {1.5}
\phi(x)=\phi^{(in)}(x)+\int D^{ret}(x-y)j(y)dy
\end   {equation} 
where
\begin {equation} 
                                                          \label {1.6}
j(x)=-g \phi^3 (x)
\end   {equation} 
The master field  $\phi(x)$ does not have matrix indexes.
Here the free scalar Boltzmannian field $\phi^{(in)}(x)$
is given by
\begin {equation} 
                                                          \label {1.7}
\phi^{(in)}(x)=\frac{1}{(2\pi)^{3/2}}\int \frac{d^{3}k}
{\sqrt{2\omega (k)}}(a^*(k)e^{ikx}+a(k)e^{-ikx}) ,
\end   {equation} 
where $\omega (k)= \sqrt{k^{2}+m^{2}}$.  It satisfies to
the Klein-Gordon equation
$$(\Box + m^2)\phi^{(in)} (x)=0$$
and it is an operator in the Boltzmannian Fock space
with relations
\begin {equation} 
                                                          \label {1.8}
a(k)a^*(k')=\delta^{(3)}(k-k')
\end   {equation} 
and vacuum $|\Omega_0),~ ~ a(k)|\Omega_0)=0$. A systematical
consideration  of the Wightman formalism  for Boltzmannian
fields is presented in \cite {AAVMin}.

Recall that if $H$ is a Hilbert space then  the Boltzmannian
Fock space $\Gamma (H)$ over $H$ is just the tensor algebra over $H$,
$ \Gamma (H)=\oplus_{n=0}^{\infty}H^{\otimes n}$ (there is no
symmetrization or antisymmetrization here). The annihilation
and creation operators are defined as
$$
a(f)f_{1}\otimes ...\otimes f_{n}=(f,f_{1})
f_{2}\otimes ...\otimes f_{n};~
a^{*}(f)f_{1}\otimes ...\otimes f_{n}=
f\otimes f_{1}\otimes ...\otimes f_{n}.
$$
One has a relation $a(f)a^{*}(h)=(f,h)$ where $(f,h)$
is an inner product in $H$. In our case $H=L^2(R^3)$
and one uses notations such as $a(f)=\int a(k)f(k)dk$.
An $n$-particle state is created from the vacuum $|\Omega_{0})
=1$ by the usual formula
$$
|k_{1},...,k_{n})=a^{*}(k_{1})...a^{*}(k_{n}) |\Omega_{0})
$$
but it is not symmetric under permutation of $k_{i}$.

Our master field $\phi(x)$ satisfies to the standard equation
for the scalar field
\begin {equation} 
                                                          \label {1.9}
(\Box + m^{2})\phi (x)=-g\phi ^3(x).
\end   {equation} 
The only difference of $\phi (x)$ from the standard scalar
quantum field is in the relation (\ref {1.8}). Let us show that
the following theorem is true.

{\bf Theorem 1.} {\it At every order of perturbation theory in the
coupling constant one has the following relation
\begin {equation} 
                                                          \label {1.10}
\lim _{N \to \infty}
\frac{1}{N^{1+\frac{k}{2}}}<0|\tr(M(x_1)...M(x_k))|0>
=(\Omega_{0}|\phi (x_{1})...\phi (x_{k})|\Omega_{0})
\end   {equation} 
where   the field $M (x)$ is defined by  (\ref {1.3})
and  $\phi (x)$ is defined by  (\ref {1.5})}.

To prove the theorem let us expand (\ref {1.3}) into the
perturbative series in the coupling constant $g$,
$$
M(x)=M^{(in)}(x)+gM^{(1)}(x) +g^{2}M^{(2)}(x)+...
$$
Then at the $n$-th order of the perturbation series
one gets a polynomial over
$M^{(in)}$ integrated with a number of propagators
$D^{ret}$,
\begin {equation} 
                                                          \label {1.11}
M^{(n)}(x)=\frac{1}{N^{n}}\int Y^{(n)}(x;y_{1},...,y_{2n+1})
M^{(in)}(y_{1})...M^{(in)}(y_{2n+1})dy_{1}...y_{2n+1}
\end   {equation} 
Here the function  $Y^{(n)}$ does not depend on $N$ and it is
a polynomial over  $D^{ret}$  and $\delta$-functions.
Now let us expand the Wightman functions  (\ref {1.4}) into the
perturbative series in the coupling constant $g$,
$$
W(x_{1},...,x_{k})=W^{(0)}(x_{1},...,x_{k})+
gW^{(1)}(x_{1},...,x_{k})+...
$$
One has
\begin {equation} 
                                                          \label {1.12}
W^{(n)}(x_{1},...,x_{k})=\frac{1}{N^{1+\frac{k}{2}}}
\sum_{|p|=n}<0|\tr(M^{(p_{1})}(x_{1})...M^{(p_{k})}(x_{k}))|0>
\end   {equation} 
where $|p|=p_{1}+...+p_{k}$. From  (\ref {1.11}) and (\ref {1.12})
one gets
\begin {equation} 
                                                           \label {1.13}
W^{(n)}(x_{1},...,x_{k})=
\sum_{|p|=n}
\int Y^{(p_{1})}(x_{1};y_{1},...,y_{2p_{1}+1})...
     Y^{(p_{k})}(x_{k};z_{1},...,z_{2p_{k}+1})
\end   {equation} 
$$
\frac{1}{N^{1+\frac{k}{2}+n}}
<0|\tr(M^{(in)}(y_{1})...M^{(in)}(y_{2p_{1}+1})....
M^{(in)}(z_{1})...M^{(in)}(z_{2p_{k}+1}))|0>dy_{1}...dz_{2p_{k}+1}
$$
To perform the limit $N \to \infty $ we will use the following
basic relation
\begin {equation} 
                                                          \label {1.B}
\lim _{N \to \infty}
\frac{1}{N^{1+\frac{k}{2}}}<0|\tr(
(M^{(in)}(y_{1}))^{p_{1}}...(M^{(in)}(y_{r}))^{p_{r}})|0>
\end   {equation} 
$$
=(\Omega_{0}|(\phi^{(in)} (y_{1}))^{p_{1}}...(\phi^{(in)} (x_{r})
)^{p_{r}}|\Omega_{0})
$$
where $k=p_{1}+...+p_{r}$.
To prove  (\ref {1.B})  one uses the Wick theorem
for the Wightman functions and 't Hooft's graphs with double lines.
One has
\begin {equation} 
                                                          \label {1.14}
<0|M^{(in)}_{ij}(x)M^{(in)}_{pq} (y)|0>=\delta_{iq}\delta_{jp}
D^{-}(x-y)
\end   {equation} 
where
$$
D^{-}(x)=\frac{1}{(2\pi)^{3}}\int e^{ikx}\theta (-k^{0})
\delta (k^{2}-m^{2})dk.
$$
According to the Wick theorem we represent  the vacuum
expectation value in the L.H.S. of (\ref {1.13}) as a sum
of 't Hooft's graphs with the propagators   (\ref {1.14}).
Then in the limit  $N \to \infty $  only non-crossing
(rainbow) graphs are nonvanished. We get the same expression
if we compute the R.H.S. of (\ref {1.13}) by using the relations
(\ref {1.8}), i.e. by using the Boltzmannian Wick theorem.
This gives the proof of the relation (\ref {1.13}).

 From (\ref {1.13}) and (\ref {1.B}) it follows that
$$
\lim _{N \to \infty}
W^{(n)}(x_{1},...,x_{k} )=
\sum_{|p|=n} \int Y^{(p_{1})}(x_{1};y_{1},...,y_{2p_{1}+1})...
 Y^{(p_{k})}(x_{k};z_{1},...,z_{2p_{1}+1})
$$
\begin {equation} 
                                                          \label {1.15}
(\Omega_{0}|\phi^{(in)} (y_{1})...\phi^{(in)} (y_{2p_{1}+1})...
\phi^{(in)} (z_{1})...\phi^{(in)} (z_{2p_{k}+1})
|\Omega_{0})dy_{1}...dz_{2p_{k}+1}
\end   {equation} 
Now if one expands equation  (\ref {1.9})  into the perturbative series
over the coupling constant $g$,
$$
\phi (x)=  \phi^{(in)} (x)+g\phi^{(1)} (x)+g^{2}\phi^{(2)} (x)
+...
$$
then one has
$$
\phi^{(n)} (x)  =
\int Y^{(n)}(x;y_{1},...,y_{2n+1})...
\phi^{(in)} (y_{1})...\phi^{(in)} (y_{2n+1})dy_{1}...dy_{2n+1}
$$
where functions $Y^{(n)}$ are the same that we have in the
expression  (\ref {1.11}) for the expansion of the matrix
field $M(x)$ into the perturbative series. Therefore the R.H.S.
of   (\ref {1.15})  is equal to
$$
\sum_{|p|=n} (\Omega_{0}|\phi^{(p_{1})} (x_{1})...
\phi^{(p_{k})} (x_{k})|\Omega_{0}),
$$
which is  the $n$-th term in the perturbative expansion of the R.H.S.
of  (\ref {1.10})  in the coupling constant.
This finishes the proof of Theorem 1.

Renormalization of the master field can be performed in the same way
as the renormalization of the ordinary Yang-Feldman
equation \cite {Dy,Kal,BD}.
The renormalized form of the Yang-Feldman equation reads
$$
\Phi(x)=\sqrt {Z}\Phi^{(in)}(x)+\int D^{ret}(x-y)
[g_{R}\Phi^{3}(y)+\delta m^{2}\Phi (y)]dy.
$$

The Boltzmannian relations   (\ref {1.8}) can be considered
as a special case of a $q$-deformed quantum field theory
(see  for example \cite {Coo} -\cite {APVV}, \cite {ALV,AAKV}
and the references therein) if we quantize a field $\phi (x)$
satisfying to equation (\ref {1.9}) assuming $q$-commutation
relations for the in-field
\begin {equation} 
                                                          \label {5.2}
a(k)a^*(k')-qa^*(k')a(k)=\delta^{(3)}(k-k')
\end   {equation} 
where $q$ is a parameter. It would be interesting to consider
in detail properties of this $q$-deformed model and to see a dependence
of renormalization constants on the parameter $q$.

\section{Gauge field}
\setcounter{equation}{0}

In this section we construct the master field for  gauge field
theory. Let us consider the Lagrangian
\begin {equation} 
                                                          \label {3.1}
L=\tr\{-\frac{1}{4}F_{\mu \nu}^{2}-\frac{1}{2\alpha}
(\partial_{\mu}A_{\mu})^{2}+\bar{c}\partial_{\mu}
\nabla_{\mu}c\}
\end   {equation} 
where  $A_{\mu}$ is the gauge field for the $SU(N)$ group,
$c$ and $\bar{c}$ are the Faddeev-Popov ghost fields  and $\alpha$
is a gauge fixing parameter.  The fields $A_{\mu}$, $c$ and $\bar{c}$
take values in the adjoint representation. Here
$$
F_{\mu \nu}=\partial_{\mu}A_{\nu}-\partial_{\nu}A_{\mu}
+\frac{g}{N^{\frac{1}{2}}}[A_{\mu},A_{\nu}],~~~
\nabla_{\mu}c=\partial_{\mu}c+\frac{g}{N^{\frac{1}{2}}}[A_{\mu},c],
$$
$g$ is the coupling constant.
Equations of motion have the form
\begin {equation} 
                                                          \label {3.2'}
\nabla_{\mu} F_{\mu \nu} +
\frac{1}{\alpha}\partial_{\nu}\partial_{\mu}A_{\mu}+
\frac{g}{N^{\frac{1}{2}}}\partial_{\nu}\bar{c} c+
\frac{g}{N^{\frac{1}{2}}}c\partial_{\nu}\bar{c}=0,
\end   {equation} 
$$
\partial_{\mu}(\nabla_{\mu}c)=0,~~~
\nabla_{\mu}(\partial_{\mu}\bar{c})=0
$$
One writes these equations in the form
\begin {equation} 
                                                          \label {3.2}
\Box   A_{\nu} -(1-  \frac{1}{\alpha})\partial_{\nu}
\partial_{\mu}A_{\mu}=J_{\nu},
\end   {equation} 
$$
\Box c=J,~~~\Box \bar{c}=\bar{J},
$$
where
$$
J_{\nu} = -\frac{g}{N^{\frac{1}{2}}}\partial_{\mu}[A_{\mu},A_{\nu}]
-\frac{g}{N^{\frac{1}{2}}}[A_{\mu},F_{\mu\nu}]-
\frac{g}{N^{\frac{1}{2}}} \partial_{\nu} \bar{c}c-
\frac{g}{N^{\frac{1}{2}}}c\partial_{\nu}\bar{c},
$$
$$
  J = -\frac{g}{N^{\frac{1}{2}}} \partial_{\mu}[A_{\mu},c],~~~
\bar{J}=- \frac{g}{N^{\frac{1}{2}}}[A_{\mu},\partial_{\mu}\bar{c}]
$$
 From  (\ref {3.2}) one gets the Yang-Feldman equations
\begin {equation} 
                                                          \label {3.3}
A_{\mu}(x)=A_{\mu}^{(in)}(x)+  \int D^{ret}_{\mu\nu}
(x-y)J_{\nu}(y)dy,
\end   {equation} 
$$c(x)=c^{(in)}(x)+  \int D^{ret}(x-y)J(y)dy,
{}~ ~\bar{c}(x)=\bar{c}^{(in)}(x)+
\int D^{ret}(x-y)\bar{J}(y)dy,
$$
where
$$
D^{ret}_{\mu\nu}(x)=(g_{\mu\nu}-(1-\alpha )\frac
{\partial_{\mu}\partial_{\nu}}{\Box})D^{ret}(x),
$$
and $g_{\mu\nu}$ is the Minkowski metric. Free in-fields
satisfy
$$
(\Box g_{\mu\nu}-(1-\frac{1}{\alpha} )
\partial_{\mu}\partial_{\nu})A_{\nu}^{(in)}(x)=0,
$$
$$
\Box  c^{(in)}(x)=0,~ ~\Box \bar{c}^{(in)}(x)=0.
$$
and they are quantized in the Fock space with vacuum $|0>$.
The vector field $A_{\mu}^{(in)}$ is a Bose field and
the ghost fields $c^{(in)},\bar{c}^{(in)}$ are Fermi
fields. Actually one assumes a gauge $\alpha =1$. In a different
gauge one has to introduce additional ghost fields.
We introduce the notation $\psi_{i}=(A_{\mu},c,\bar {c})$
for the multiplet of  gauge and ghost fields.
The $U(N)$-invariant
Wightman functions  are defined as
\begin {equation} 
                                                          \label {3.4}
W(x_{1},...,x_{k})=
\frac{1}{N^{1+\frac{k}{2}}}<0|tr(\psi_{i_{1}}(x_{1})...
\psi_{i_{k}}(x_{k}))|0>.
\end   {equation} 
We will show that the limit of functions (\ref {3.4}) when
$N\to \infty$ can be expressed in terms of the master fields.
The master field for the gauge field $A_{\mu}(x)$ we denote
$B_{\mu}(x)$ and the master fields for the ghost fields
$c(x),\bar {c}(x)$ will be denoted $\eta(x),\bar {\eta}(x)$ .
The master fields  satisfy to equations
$$
D_{\mu} {\cal F}_{\mu \nu} +
\frac{1}{\alpha}\partial_{\nu}\partial_{\mu}B_{\mu}+
g\partial_{\nu}\bar{\eta} \eta +g\eta\partial_{\nu}\bar{\eta}=0,
$$
\begin {equation} 
                                                          \label {3.5'}
\partial_{\mu}(D_{\mu}\eta)=0,~~~
D_{\mu}(\partial_{\mu}\bar{\eta})=0
\end   {equation} 
where
\begin {equation} 
                                                          \label {3.0}
{\cal F}_{\mu \nu}=\partial_{\mu}B_{\nu}-\partial_{\nu}B_{\mu}
+g[B_{\mu},B_{\nu}],~ ~  ~
D_{\mu}\eta=\partial_{\mu}\eta+g[B_{\mu},\eta].
\end   {equation} 
These equations have the form of the Yang-Mills equations
 (\ref {3.2}) however the master fields
$B_{\mu}$, $\eta$, $\bar {\eta}$ do not have  matrix indexes
and they do not take values in a finite dimensional Lie algebra.
The gauge group for the field $B_{\mu}$ is an infinite dimensional group
of unitary operators in the Boltzmannian Fock space.
Equations (\ref {3.5'}) in terms of currents read
\begin {equation} 
                                                          \label {3.5}
\Box   B_{\nu} -(1-  \frac{1}{\alpha})\partial_{\nu}
\partial_{\mu}A_{\mu}=j_{\nu},
\end   {equation} 
$$
\Box \eta=j,~ ~~\Box \bar{\eta}=\bar{j},
$$
where
$$
j_{\nu} = -g\partial_{\mu}[B_{\mu},B_{\nu}]- g[B_{\mu},
{\cal F}_{\mu \nu}] - g\partial_{\nu}\bar{\eta}\eta
-g\eta\partial_{\nu}\bar{\eta},
$$
$$
j=-g \partial_{\mu}[B_{\mu},\eta],~~ ~\bar{j}=
- g[B_{\mu},\partial_{\mu}\bar{\eta}].
$$

We define the master fields by using the Yang-Feldman equations
\begin {equation} 
                                                          \label {3.6}
B_{\mu}(x)=B_{\mu}^{(in)}(x)+  \int D^{ret}_{\mu\nu}
(x-y)j_{\nu}(y)dy,
\end   {equation} 
$$
\eta(x)=\eta^{(in)}(x)+  \int D^{ret}(x-y)j(y)dy,
{}~ ~\bar{\eta}(x)=\bar{\eta}^{(in)}(x)+
\int D^{ret}(x-y)\bar{j}(y)dy,
$$
The in-master fields are quantized in the Boltzmannian Fock space. For
the master gauge field we have
\begin {equation} 
                                                          \label {3.7}
B_{\mu}^{(in)}(x)=\frac{1}{(2\pi)^{3/2}}\int \frac{d^{3}k}
{\sqrt{2|k|}}\sum _{\lambda =1}^{4}\epsilon ^{(\lambda)}
_{\mu}(k)[a^{(\lambda )*}(k)e^{ikx}+a^{(\lambda )}(k)e^{-ikx}) ,
\end   {equation} 
where $\epsilon ^{(\lambda)}_{\mu}(k)$  are polarization vectors
and annihilation and creation operators satisfy
\begin {equation} 
                                                          \label {3.8}
a^{(\lambda )}(k)a^{(\lambda ')*}(k')=
g^{\lambda \lambda '}\delta ^{(3)}(k-k'),
\end   {equation} 
The expression (\ref {3.7}) for the field $B_{\mu}(x)$ looks like
an expression for the photon field. However because of relations
(\ref {3.8})  the commutator  $[B_{\mu}(x),B_{\nu}(x)]$   does not
vanish and it  permits us to develope a gauge theory for the field
$B_{\mu}(x)$ with a non-Abelian gauge symmetry.

We quantize the master ghost fields  in the Boltzmannian Fock space
with indefinite metric
\begin {equation} 
                                                          \label {3.10}
\eta^{(in)}(x)=\frac{1}{(2\pi)^{3/2}}\int \frac{d^{3}k}
{\sqrt{2|k|}}(\gamma^*(k)e^{ikx}+\beta (k)e^{-ikx}) ,
\end   {equation} 
$$
\bar{\eta}^{(in)}(x)=\frac{1}{(2\pi)^{3/2}}\int \frac{d^{3}k}
{\sqrt{2|k|}}(\beta ^*(k)e^{ikx}+\gamma(k)e^{-ikx}) ,
$$
where creation and annihilation operators satisfy
$$
\gamma (k)\gamma ^{*}(k')=\delta ^{(3)}(k-k'),
$$
\begin {equation} 
                                                          \label {3.12}
\beta (k)\beta ^{*}(k')=-\delta ^{(3)}(k-k').
\end   {equation} 
We also assume that the product of any annihilation operator
with a creation operator of a different type
always is equal to zero, i.e.
$$
\gamma (k) \beta ^{*}(k')=\beta (k)\gamma ^{*}(k')=
a^{(\lambda )}(k)\gamma ^{*}(k)=0,$$
\begin {equation} 
                                                          \label {3.14}
a^{(\lambda )}(k)\beta ^*(k')=
\gamma (k)a^{(\lambda )*}(k')=\beta (k)a^{(\lambda )*}(k')=0.
\end   {equation} 
The Boltzmannian Fock vacuum satisfies
\begin {equation} 
                                                          \label {3.15}
\gamma (k)|\Omega _{0})=\beta (k)|\Omega _{0})=
a^{(\lambda )}(k)|\Omega _{0})=0.
\end   {equation} 
Let us denote $\chi _{i}=(B_{\mu}, \eta, \bar{\eta})$
the multiplet of the
master fields. The following theorem is true.

{\bf Theorem 2.} {\it At every order of perturbation
theory in the
coupling constant one has the following relation
\begin {equation} 
                                                          \label {3.16}
\lim _{N \to \infty}
\frac{1}{N^{1+\frac{k}{2}}}<0|\tr(\psi _{i_{1}}(x_1)...
\psi _{i_{k}}(x_k))|0>
=(\Omega_{0}|\chi _{i_{1}} (x_{1})...
\chi _{i_{k}}(x_{k})|\Omega_{0})
\end   {equation} 
where   the fields $A_{\mu} (x), c(x)$  and
$\bar {c}(x)$ are defined by
(\ref {3.3}) and $B_{\mu} (x), \eta(x)$  and $\bar {\eta}(x)$
are defined by  (\ref {3.6})}.

The proof of Theorem 2 is analogous to the proof of Theorem 1.
We get relations (\ref {3.12}) for master fields by taking
into account the wrong statistics of the ghost fields.

\section{Operator BRST-transformations}
\setcounter{equation}{0}

The set of planar diagrams is invariant under gauge transformations.
Therefore the theory of the master fields $B_{\mu}$, $\eta$
and $\bar{\eta}$  also should be gauge invariant.
Actually a more general fact is true.
Let be given a system of equations
\begin {equation} 
                                                          \label {4.1}
D_{\mu} {\cal F}_{\mu \nu}+
\frac{1}{\alpha}\partial_{\nu}\partial_{\mu}B_{\mu}+
\partial_{\nu}\bar{\eta} \eta +\eta\partial_{\nu}\bar{\eta}=0,
\end   {equation} 
$$
\partial_{\mu}(D_{\mu}\eta)=0,~~~
D_{\mu}(\partial_{\mu}\bar{\eta})=0
$$
where
\begin {equation} 
                                                          \label {4.4}
{\cal F}_{\mu \nu}=\partial_{\mu}B_{\nu}-\partial_{\nu}B_{\mu}
+[B_{\mu},B_{\nu}];~ ~
D_{\mu}\eta=\partial_{\mu}\eta+[B_{\mu},\eta],
\end   {equation} 
and  $B_{\mu}(x)$, $\eta(x)$ and $\bar{\eta}(x)$ are abstract operator
valued functions on space-time taking values in an associative algebra
not necessary being the master fields. We do not assume any relations
like the Hermitian conjugation or anticommutativity for $\eta$ and
$\bar{\eta}$. Let us define an operator BRST-transformations
\begin {equation} 
                                                          \label {4.7}
\delta B_{\mu}=D_{\mu}\eta \epsilon ,
\end   {equation} 
$$
\delta \eta=\eta ^{2}\epsilon,
$$
$$
\delta \bar {\eta}=-\frac{1}{\alpha}\partial_{\mu}B_{\mu} \epsilon,
$$
where $\epsilon$ is a constant infinitesimal parameter such that
\begin {equation} 
                                                          \label {4.8}
\eta \epsilon+\epsilon\eta =0, ~~
\bar{\eta} \epsilon+\epsilon\bar {\eta} =0.
\end   {equation} 
The following proposition is true.

{\bf Proposition.} {\it If conditions
(\ref {4.8}) are satisfied then equations
(\ref {4.1}) are invariant under the operator
BRST-transformations (\ref {4.7}).}

To prove the proposition one  performs straightforward
computations and uses relations
\begin {equation} 
                                                          \label {4.9}
\delta {\cal F}_{\mu \nu}=[{\cal F}_{\mu \nu},\eta]\epsilon,~~
\delta (D_{\mu} {\cal F}_{\mu \nu})=
[D_{\mu} {\cal F}_{\mu \nu},\eta]\epsilon ,
\end   {equation} 
$$
D_{\mu}D_{\nu} {\cal F}_{\mu \nu}=
-\frac{1}{2}[{\cal F}_{\mu \nu},{\cal F}_{\mu \nu}]=0,~~
\delta (D_{\mu}\eta)=0.
$$
The master field equations (\ref {3.5'}) are a special
case of general operator
equations (\ref {4.1}). Therefore the BRST-invariance
of the  master field equations (\ref {3.5'}) follows  from the proposition.

\section {Concluding Remarks}
\setcounter{equation}{0}

We have constructed in this paper the master fields
describing the large N limit
for matrix scalar, gauge and fermionic fields. Our method  based on the
Yang-Feldman equation is a general one and can
be applied to any matrix model.
It is unusual in modern quantum field theory that
to solve a problem one has
to work in the  physical Minkowski space-time instead
of the usually adopted Euclidean
formulation. Note also that the formalism of the Wightman
correlation functions is more convenient for
the consideration of the large N limit then the usually
used formalism of Feynman  correlation  functions
and interaction representation.
We can introduce a new chronological product $T_{B}$ for the
Boltzmannian master fields such that
$$
(\Omega_{0}|T_{B}(\phi^{(in)} (y_{1})...
\phi^{(in)} (x_{k}))|\Omega_{0})=
$$
$$
\lim _{N \to \infty}
\frac{1}{N^{1+\frac{k}{2}}}<0|T\tr(
(M^{(in)}(y_{1})...M^{(in)}(y_{r}))|0>
$$
Here $T$ is the ordinary chronological
ordering. An explicit definition of
$T_{B}$ is rather involved. It is similar to an
explicit realization of the canonical momenta for the free
Euclidean master field  \cite {AAV}.
A simple definition  is found for half-planar diagrams  \cite {AAV,AZ}.
The formalism  of Green functions and functional integrals is convenient
in quantum field theory when we deal with Bose or Fermi statistics, i.e.
with a simple operator field algebra with commuting or anticommuting
generators. However in the large $N$ limit we meet a more complicated
operator field algebra and here it is more convenient to work with the
Wightman correlators.

Relations (\ref {1.8})  have been considered by
Greenberg \cite {Gre} as an example of infinite statistics by
Doplicher, Haag and Roberts \cite {DHR}.

There are many interesting problems which deserve a further study.
A relation of the present approach with the Haan \cite {Haan} Euclidean
approach and with the Greensite and Halpern \cite {GH} Euclidean stochastic
approach deserves  a study.  Douglas \cite {Doug3} has discussed the
stochastic formulation of \cite {GH} in terms of free variables.

It would be interesting to investigate the BRST-formalism for the master
gauge field presented in this paper and to see in detail  gauge  invariance
in the line of consideration of  the ordinary gauge theory  \cite {Jap}.
One can get a more general gauge invariant theory if instead (\ref {3.8})
one considers $q$-deformed commutation relations for $in$-fields:
$$
a^{(\lambda )}(k)a^{(\lambda ')*}(k')-qa^{(\lambda ')*}(k')
a^{(\lambda )}(k)=
g^{\lambda \lambda '}\delta ^{(3)}(k-k'),
$$

Equations of motion for the master field
$D_{\mu} {\cal F}_{\mu \nu}=0$ are invariant under gauge transformations
$B_{\mu}\to B_{\mu}^{\omega}=\omega^{*}B_{\mu}\omega +
\frac{1}{g} \omega^{*} \partial_{\mu}\omega
$ where $\omega=\omega (x)$ is an element of the group $U({\cal F})$
of unitary operators in the Boltzmannian Fock space  ${\cal F}$.
Properties of gauge theory with an infinite dimensional gauge group
are different from a familiar gauge theory with a finite-dimensional
Lie gauge group. In particular, it is known that all homotopic groups
of the group $U(H)$ of unitary operators in a Hilbert space $H$ are trivial,
$\pi _{k}(U(H))=0$ for
$k=0,1,...$ \cite {Sch}.
Note that for the group $U(\infty)$ one has $\pi _{2k-1}(U(\infty))=Z$ and
$\pi _{2k}(U(\infty))=0$ (R.Bott). An approach to gauge theory adapted
for a consideration of an infinite dimensional gauge group has been
developed in \cite {CFV}. An investigation of
topological properties of the infinite-dimensional gauge group for the
master field can be useful  for the problem of quark confinement in QCD in
the large $N$ limit.

Another interesting question is about an analogue of the
Makeenko-Migdal loop equation \cite {MM}.
In the present approach we have the following relation
$$
W(C)=\lim _{N \to \infty}\frac{1}{N}
<0|\tr(P\exp\{\frac{g}{N^{\frac{1}{2}}}
\int _{C}A_{\mu}dx^{\mu}\})|0>=
$$
$$
(\Omega _{0}|P\exp\{g \int _{C}
B_{\mu}dx^{\mu}\}|\Omega _{0})
$$
Here $P$ is the path ordering operator. Since we deal with the
Wightman correlation functions we have the following equation
$$
\Delta _{L}W(C)=0,
$$
where $\Delta _{L}$ is the generalized Levy Laplacian,
see  \cite {AV,AGV}. Here we ignore the ghost fields. This equation is
linear in contrast to the Makeenko-Migdal loop equation.

In this note we have considered the equations of motion for the master
field. For the construction of an appropriate action we can
use the Adler  formalism.  Adler \cite {Ad} has proposed a
generalization of quantum mechanics in which a dynamics is
formulated on a manifold with non-commuting coordinates,
which act as operators on an underlying Hilbert space.
In this formalism one can make an operator gauge transformation.
If we apply this approach to our master gauge field
then we can take the Boltzmannian Fock space
as the underlying Hilbert space and consider the following Lagrangian
$$
L=\mbox{Tr}\{-\frac{1}{4}{\cal F}_{\mu \nu}{\cal F}_{\mu \nu}-
\frac{1}{2\alpha}\partial_{\mu}B_{\mu}\partial_{\nu}B_{\nu}
+\bar{c}\partial_{\mu}D_{\mu}c\}
$$
Here one assumes a $Z_{2}$-grading in the Boltzmannian
Fock space and a trace operator $\mbox{Tr}{\cal O}$ for an operator
${\cal O}$ is defined as \cite {Ad} $\mbox{Tr}{\cal O}=Re
\Sigma_{n}(n|(-1)^{F}{\cal O}|n)$  where $Re$ is the real
part, $\{ |n)\}$ is a complete set of states and
the Witten index $(-1)^{F}$ counts fermion number
modulo two. Then one can make operator variations of the
master fields and to get equations (\ref {3.5'}).

We have constructed the master field
however still there is a problem of an analytical investigation
of correlation functions of the master field.
Just to this problem one reduces
the problem of quantitative analysis of the planar approximation
for QCD.
$$~$$
$$~$$
{\bf ACKNOWLEDGMENT}
$$~$$
The authors are grateful to D.Gross and A.Migdal for
stimulating discussions of the problem of construction
of the master field.
This work is supported in part by International Science Foundation
grant M1L000. I.A. is supported
in part by Russian Foundation for Fundamental
Research grant 93-011-147.
I.V. is supported in part by Russian Foundation for Fundamental
Research grant 93-011-140.
$$~$$
\end{document}